# Aging as a process of accumulation of Misrepairs

Jicun Wang-Michelitsch[1]*, Thomas M Michelitsch[2]

[1]Department of Medicine, Addenbrooke's Hospital, University of Cambridge, UK (Work address until 2007)

[2] Institut Jean le Rond d'Alembert (Paris 6), CNRS UMR 7190 Paris, France

**Abstract**

We recently introduced Misrepair-accumulation theory as an interpretation of aging mechanism. For better understanding this theory, we discuss here in more details the new concept of Misrepair and the concept of accumulation of Misrepairs. **I.** Aging takes place uniquely on the systems that have a well-defined structure: an organization of its sub-structures. Aging of a non-living system is a result of accumulation of injuries (damage) of its structure. **II.** A generalized concept of Misrepair is important for reaching to a unified understanding of aging changes. In the situation of a severe injury, incorrect repair is a rapid way of repair that is essential for maintaining the structural integrity of an organism. Of a similar mechanism to that in "Misrepair of DNA", the term of "Misrepair" can be used for describing all kinds of incorrect repairs in different types of living structures, including molecules, cells and tissues. A new concept of Misrepair is therefore proposed, and it is defined as incorrect reconstruction of an injured living structure. Misrepair mechanism is beneficial for the survival of an organism, and it is essential for the survival of a species. **III.** Alteration of structure made by Misrepair is irreversible; therefore Misrepairs accumulate and disorganize gradually a structure, which appears as aging of it. Accumulation of Misrepairs takes place on molecular, cellular and tissue levels, respectively. On tissue level, it appears as disordering of cells and extracellular matrixes (ECMs). On cellular level, it appears as deformation of cytoskeleton and change of cell shape. On genome DNA, it appears as accumulation of DNA mutations and alteration of DNA sequence. An essential change in aging of a multi-cellular organism is an irreversible change of the spatial relationship between cells/ECMs in a tissue. In conclusion, aging of an organism is a result of accumulation of Misrepairs on tissue level.

**Keywords**

Aging, Misrepair, accumulation of Misrepairs, injury, living structures, incorrect repair, structural integrity, survival, species' survival, alteration of structure, irreversible change, aging of a cell, aging of a molecule, aging of an organism, and tissue level

Aging is nowadays a popular topic. Some biological theories have been proposed for interpreting aging mechanism; however they all are untenable on certain aspects. A key step for uncovering aging secret is actually to find out a change, which takes place often in an organism and which can remain permanently. Along this direction, we found out that an incorrect repair like that in scar formation is such a change. Incorrect repair can take place when an injury of a cell or a tissue is severe; and the structural alteration by incorrect repair can remain permanently. On this basis, we proposed a generalized concept of Misrepair in the Misrepair-accumulation theory, for describing all manners of incorrect repairs (Wang et al, 2009). Misrepair is a result of "SOS" repair for increasing the surviving chance of an organism by maintaining the structural integrity. However, Misrepairs are irreversible and accumulating, leading to disordering and aging of a cell or a tissue. For better understanding this theory, in the present paper, we will make a detailed discussion on the new concept of Misrepair and on the concept of accumulation of Misrepairs. Our discussion will cover the following issues:

I. A generalized concept of Misrepair

    1.1 Incorrect repair as a strategy of repair
    1.2 A generalized concept of Misrepair
    1.3 Examples of Misrepairs

II. Aging as a process of accumulation of Misrepairs

    2.1 Accumulation of Misrepairs and deformation of a structure
    2.2 Accumulation of Misrepairs on molecular level
    2.3 Accumulation of Misrepairs on cellular level
    2.4 Accumulation of Misrepairs on tissue level
    2.5 Aging of an organism: on tissue level

III. Conclusion: Misrepair-accumulation theory

## I. A generalized concept of Misrepair

Aging is a universal phenomenon in nature; however it takes place only in the systems that have a well-defined structure. A colony of bacteria has no aging, because the bacteria in a colony are not in a functional organization. Differently, an animal and a plant have a well-defined structure, namely, a special organization of its sub-structures, including organs, tissues, cells and molecules. The property of a structure is determined by the organization of its sub-structures; and a change on this organization will lead to a change of its property and functionality. Aging of an organism is in fact a gradual alteration of its structure accompanied with loss of functionality.

### 1.1 Incorrect repair as a strategy of repair

Injuries, as a consequence of steady damage-exposure, are unavoidable for a living organism. An injury (an original damage) is in fact a defect of the structure of a molecule, a cell or a tissue. By destroying the structural integrity, an injury, if not repaired, will lead to failure of the structure and death of the whole organism. For example, if a wound on skin is not repaired, one can die from bleeding and infection. Repair of an injury is essential for maintaining the structural integrity and the functionality of a living structure for the survival of the organism. For repairing an injury of a tissue, a great deal of molecules and cells need to be produced and activated. A series of biochemical reactions need to take place successively and accurately, including expression of genes, production and secretion of proteins and small molecules, activation of enzymes, transportation of substance, and migration of immune cells. It is time-consuming for an organism to accomplish all of these reactions successively; thus repair of an injury needs time to be achieved.

Although the repairing process of an injury in a tissue and in a cell is complex, the result of repair is one of the followings: complete repair, failure of repair, and incorrect repair (Table 1). When a lesion is small, full repair can be achieved and the injured structure can be completely restored. If a lesion is too large, complete repair will be impossible to achieve, and an organism may die from failure of repair. A strategy for increasing the surviving chance of an organism from a severe injury is incorrect repair, a repair with altered materials or in altered remodeling, for closing the defect quickly. Scar formation on the skin and in other organs tells us that such a compromising strategy of repair exists in living beings. Small but repeated injuries are also difficult to be completely repaired, since the repairing processes are disturbed by newly arrived injuries. Repeated constriction/dilatation of arterial walls, repeated smoking, repeated UV light-exposures, and chronic inflammations are the origins of repeated injuries of tissues and cells. For such injuries, incorrect repair is unavoidable. Likely, for an injury with un-degradable substance from dead cells, isolating the substance in a capsule is a way of repair for rebuilding a functional structure of local tissue. In these last three situations, quick but incorrect repairs, no matter in what a manner, have to be made to maintain structural integrity for preventing death of the organism (Table 1).

Different tissues have different potentials of repair, and some tissues are regenerable and some are un-regenerable. Regenerable tissues are the tissues, in which the cells can be reproduced and dead cells can be replaced by new cells. Epithelium, endothelium, hepatic tissue, and connective tissue are all regenerable tissues. In these tissues, small injuries can be fully repaired and restored; but severe injuries, repeated injuries, and the injuries with un-degradable substances have to be repaired incompletely. Differently, in un-regenerable tissues like nerve tissue and skeleton muscular tissue, an injury with death of cells, has to be repaired with other types of cells and/or extracellular matrixes (ECMs). For example, dead neuron cells will be replaced by glial cells, and dead muscular cells will be replaced by fibroblasts and collagen fibers (Table 1). On cellular level, although a cell can regenerate its sub-structures, complete repair could be only possibly achieved when the injury is not severe. For severe injuries and repeated injuries, incorrect repair is the way of repair for survival of the cell.

**Table 1.** Results of repairs in tissues

| Injuries | Results of repair | |
| --- | --- | --- |
| | In regenerable tissues | In un-regenerable tissues |
| Small injury with death of cells | Complete repair | Incorrect repair |
| Large injury | Incorrect repair | Incorrect repair |
| Repeated injuries | Incorrect repair | Incorrect repair |
| Injury with un-degradable substance | Incorrect repair | Incorrect repair |

### 1.2 A generalized concept of Misrepair

Incorrect repair is one of the strategies of repair in cells and in tissues. Such "SOS" repair can save our life many times a day. Misrepair of DNA is an incorrect repair on molecular level, which is referred to the Mis-match repair of DNA. It has been realized that Misrepair of DNA is a survival strategy of a cell in situations of severe DNA injuries (Little, 2004). Misrepair of DNA is an important source of DNA mutations in somatic cells and in tumor development (Suh, 2006). Since Misrepair of DNA has a similar mechanism to that of incorrect repair in a tissue, we propose a generalized concept of Misrepair for describing all types of incorrect repairs. The new concept of Misrepair is therefore defined as: *incorrect reconstruction of an injured living structure*. This concept is applicable to all types of living structures including molecules, cells, tissues and organs. A generalized concept of Misrepair improves our understanding of life on three points: **A**. Misrepair mechanism exists in a living being; **B**. Misrepairs are a kind of compromises of an organism in struggling with destructive environment; and **C**. Misrepair is the direct force in altering the living structures as a result of injuries. In this paper, the term of Misrepair is used for describing not only the process but also the result of Misrepair, which appears as an alteration of a structure. Actually some scientists have hypothesized the existence of Misrepair mechanism on cellular level and tissue level. For example, Tobias has suggested a Repair-Misrepair mathematical model for analyzing the survival chance of cells exposed to radiation (Tobias, 1985). Kondo has proposed a tissue-Misrepair hypothesis on explaining the radiation-associated carcinogenesis (Kondo, 1991). Unfortunately in these studies, the term of Misrepair was not clearly defined.

With a clear definition, a Misrepair is observable and detectable. Alteration of a structure made by Misrepair exhibits in different ways: **A.** a change on the amount of the sub-structures, **B.** a change on the types of the sub-structures, and/or **C.** a change on the spatial relationship between the sub-structures. It is essential to distinguish the structural change in an injury from that in a Misrepair. Like that between a wound and a scar, an injury is a change of a structure before repair, which is caused by physical or chemical damage. Differently, a Misrepair is an alteration of a structure as a result of repair. An injury appears as a defect of a structure, whereas a Misrepair appears as a change of the organization of sub-structures (Figure 1). In scientific literatures, the term of injury and the term of damage have not been clearly defined, and this is one reason why we cannot understand aging mechanism. For avoiding misunderstanding, in our theory, the term of damage is referred to a physical or chemical impact to a structure, which causes an injury (defect) of the structure. An injury is the direct effect of damage, and Misrepair is the biological response to the injury. Damage is the force that triggers the occurrence of injuries and then Misrepairs; therefore damage is the cause, whereas Misrepair is the final effect.

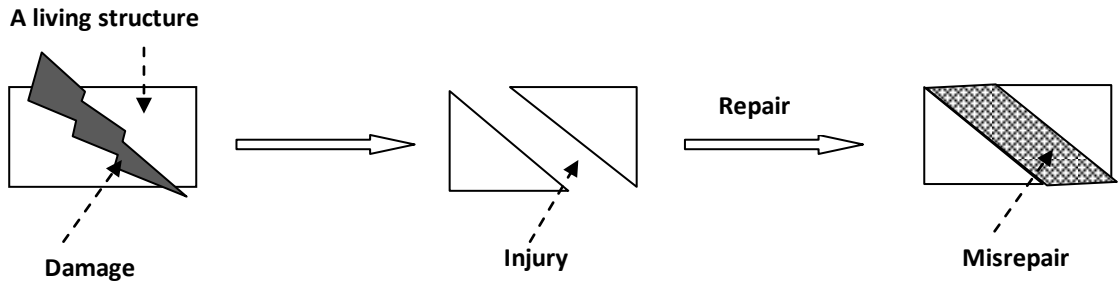

**Figure 1. Misrepair: an incorrect reconstruction of an injured living structure**

For understanding aging, it is important to distinguish three concepts: damage, injury and Misrepair. Damage is referred to a physical or chemical impact to a structure, which causes an injury of the structure (**Damage**). An injury is a defect of a structure before repair (**Injury**). Misrepair is an incorrect reconstruction of an injured structure, resulting in alternation of the structure (**Misrepair**). An injury is the direct effect of damage, and Misrepair is the biological response to the injury. Damage is the force that triggers the occurrence of injuries and then Misrepairs; therefore damage is the cause, whereas Misrepair is the final effect.

Misrepair is neither a fault nor a mistake of an organism due to the limitation of repair/maintenance, but a result of active repair, which is essential for the survival of a living being. An organism having normal repair/maintenance system DOES NOT SLOW DOWN but even PROMOTES Misrepairs when it is necessary. Without Misrepairs, an individual was unable to survive till reproduction age; therefore Misrepair mechanism is essential for the survival of a species. Species' survival is simply a result of the survivals of its individuals till reproduction age.

## 1.3 Examples of Misrepairs

A Misrepair is actually a tiny scar; however it may appear quite different in different structures and in different situations. Although we have the techniques for high-resolution imaging such as electron micro-graphing and con-focal imaging, most Misrepairs in molecules, cells, and tissues, are too small to detect. A visible pathological change in an organ is often a result of multiple times of Misrepairs. Here we give some examples of Misrepairs: **A**. **on tissue level:** scar formation, altered remodeling of cells and/or ECMs, esophagus intestinal metaplasia, and formation of Langhans giant cells; **B**. **on cellular level:** altered remodeling of cytoskeleton; and **C. on molecular level:** Misrepair of DNA (Table 2).

**Table 2. Examples of Misrepairs**

| Tissue level | Cellular level | Molecular level |
|---|---|---|
| Scar formation | Altered remodeling of cytoskeleton | Misrepair of DNA |
| Altered remodeling of cells/ECMs | | |
| Esophagus intestinal Metaplasia | | |
| Formation of Langhans giant cells | | |

Scar formation on the skin is a result of repair of epidermis and derma by connective tissue. Basement membrane in epidermis has low potential of regeneration; and a defect of basement membrane in a deep wound has to be closed by fibroblasts at first and by collagen fibers later. Scaring is a repair for sealing, and it is a powerful evidence for the existence of Misrepair mechanism. Scaring takes place often in the skin and in connective tissues; however a process of scaring can be too small to be detected. Apart from scar formation, altered remodeling of cells/ECMs has been observed in other manners. For example, a defective remodeling of ECMs, appearing as a dense and poor organization of ECMs, is observed in the myocardial tissues of the patients with Chagas' disease. Chagas' disease is a disease that is caused by infection of protozoan parasite *Trypanosoma cruzi* (Spinale, 2007; Rossi, 2001). An extensive dendritic remodeling of retina cells was observed during normal aging of human retina (Eliasieh, 2007). An altered interaction of cells through neurite sprouting and abnormal axonal projections of cone photoreceptors was seen in age-related macular degeneration (AMD) of human retina (Pow, 2007).

Esophagus intestinal metaplasia is also a result of Misrepair, and the injured esophagus mucosa is repaired by intestinal epithelium. Metaplasia is an adaptive change, since the intestinal epithelium makes local esophagus mucosa more robust to certain types of damage (Stein, 1993). Langhans giant cells are huge cells that develop from fusion of multiple macrophages. Such giant cells can be observed in granulomatous diseases such as lung transbronchial and lymph node sarcoidosis. A characteristic of a Langhans giant cell is the horseshoe-shaped distribution of multi-nuclei in cell periphery and the deposition of necrotic tissue (Caseous necrosis) in cell centre (Lay, 2007). Fusion of macrophages is a result of repair, in which the un-degradable substance of dead cells is isolated in the fused macrophages. Deposition of a Langhans giant cell has altered the local structure of a tissue, and it is a Misrepair of the tissue.

Altered remodeling of cytoskeleton is an example of Misrepair on cellular level. Cytoskeleton is the architecture of a cell, and it is composed of three types of filaments: microfilament, microtubule, and intermediate filaments. The structure of cytoskeleton is important for maintaining cell shape, nuclear shape and a functional organization of organelles. Cytoskeleton filaments compose also the pathways for substance transportation via cytoplasm or via filament cables. The organization of cytoskeleton filaments are in a dynamic modeling-remodeling state for adapting to cell functions. An altered and irreversible remodeling of cytoskeleton filaments may take place when part of the cytoskeleton structure is severely injured. For example, an altered remodeling of cytoskeleton proteins including Tintin protein, desmin filaments, and microtubules, have been found to contribute to the lengthening and malfunction of myocytes in chronic pressure-overloading and congestive heart failure (Wang, 1999). On molecular level, Misrepair of DNA is the example. The techniques for DNA sequencing make us able to detect a change on molecular level: a DNA change on "one" base-pair! In somatic cells, Misrepair of DNA is the main source of DNA mutations (Suh, 2006).

## II. Aging as a process of accumulation of Misrepairs

Misrepairs make us survive everyday; however they alter the structures of our cells and tissues and reduce their functionality. Like that in scar formation, a Misrepair is an irreversible change as a final result of repair. Being unavoidable and irreversible, Misrepairs will remain and accumulate. Accumulation of Misrepairs will gradually deform the structure of a molecule, a cell or a tissue, making aging of it.

### 2.1 Accumulation of Misrepairs and deformation of a structure

As discussed above, distinguishing a Misrepair from an injury is a key step for understanding aging mechanism. Damage (faults)-accumulation theory proposed that: it is due to the limitation of maintenance and repair that some faults can take place and accumulate, and these faults are the origin of aging (Kirkwood, 2005). However, a fault is in fact an intrinsic damage, and it results in a defect of the structure of a molecule/cell/tissue; thus it cannot remain if not repaired. For example, when a cell produces un-functional proteins by a fault and when the proteins are not removable, deposition of the proteins will result in a structural defect of the cell. By interrupting intracellular substance-transportation and information-transmission, the deposit will cause cell death. Therefore, "unrepaired" or "untreated" faults cannot remain and accumulate in a living organism. Differently, Misrepairs are the results of repair and they can accumulate. The accumulation of alteration of a structure after each time of Misrepair will gradual disorganize the structure, which appears as aging of it. For example, the accumulation of scars on the skin is part of aging of the skin.

The process of gradual deformation of a living structure by accumulation of Misrepairs is similar to the gradual alteration of a cloth textile, which has been repaired many times by different types of textile materials. Repair of a defect of a cloth textile by an altered type of textile is a kind of Misrepair. After several times of such Misrepairs in a local part of the textile, such as M1, M2, M3, M4 M5, M6, and M7, the different repairing materials will overlap each other and gradually deform this part of textile (Figure 2).

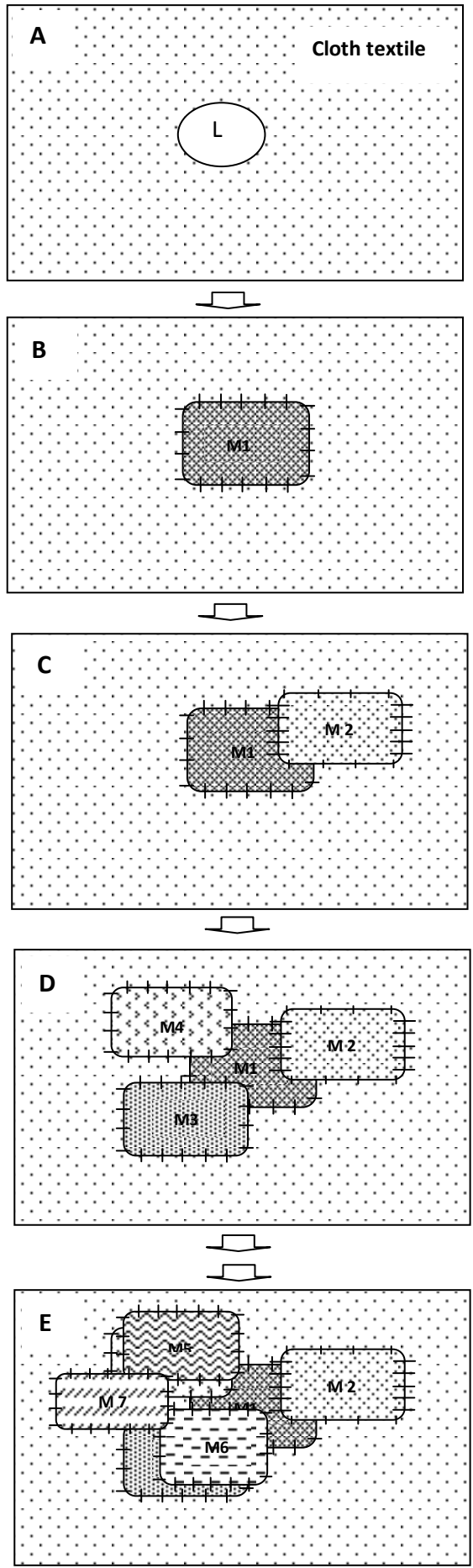

**Figure 2. Schematic representation of structure-deformation by accumulation of Misrepairs**

The process of gradual deformation of a living structure by accumulation of Misrepairs is similar to the gradual change of a cloth textile, which has been repaired many times by different types of textile materials. Repair of a lesion (defect) of a cloth textile by an altered type of textile is a kind of Misrepair. For example, a textile (**A**) is firstly repaired for closing the lesion (**L**) by Misrepair 1 (**M1** in **B**), then by Misrepair 2 (**M2** in **C**), and then by M3 (**M3** in **D**). After several times of such Misrepairs in this part of textile, the repairing materials of **M1, M2, M3, M4 M5, M6,** and **M7** will overlap each other (**M1 + M2 + M3 + M4 + M5 + M6 + M7)** and deform gradually this part of textile.

Accumulation of Misrepairs leads to a gradual disordering of a living structure and gradual reduction of its functionality. Aging of a structure is not essentially a reduction of number of its sub-structures, not essentially a change of types of its sub-structures, and not essentially a deposition of "waste", but essentially and sufficiently an accumulation of alterations of the structure by Misrepairs. Aging of an organism is a result of accumulation of Misrepairs of its structure. For non-living structures that cannot be repaired, such as architecting materials, some injuries can remain and accumulate. Therefore "aging of a material" is a result of accumulation of injuries. In fact, there are also non-living structures in an organism. Hairs, nails, the lens, and intervertebral discs are tissues that have lost biological functionality and they are dead structures. Aging of these tissues are not due to Misrepairs but due to injuries.

Animals and plants are multi-cellular beings, and a multi-cellular being is composed of five levels of sub-structures: molecules, organelles, cells, tissues, and organs. Molecules and cells are the units of repairing materials. Thus, from the aspect of repair, these sub-structures are on three levels: molecular, cellular and tissue. DNAs are on molecular level, and they are repaired by small molecules such as nuclide bases. Organelles and cells are on cellular level, and they are repaired by small molecules and big molecules such as proteins. Tissues, organs and the whole organism are on tissue level, and they are repaired by cells and ECMs. Misrepairs and accumulation of Misrepairs take place on these three levels respectively.

## 2.2 Accumulation of Misrepairs on molecular level

Accumulation of Misrepairs (mutations) of DNA is on molecular level. In somatic cells, DNA changes include chromosome changes and point DNA mutations (called DNA mutations or gene mutations). Chromosome changes are mostly fatal for a somatic cell, thus they cannot remain and accumulate in cells. Differently, DNA mutations are often silent or mild to a cell, thus they can "survive" in cells. Therefore, DNA mutations are the main type of DNA changes that accumulate in DNAs and in cells. Misrepair is the main source of DNA mutations in somatic cells (Suh, 2006). Since the surviving rate of a cell through Misrepair of DNA is low, accumulation of DNA mutations can take place only possibly in the cells that can proliferate. Accumulation of DNA mutations needs to proceed over many generations of cells; thus it is a slow process (Wang-Michelitsch, 2015). Accumulation of DNA mutations alters gradually the structure and the functionality of a DNA, leading to aging of the DNA. Therefore, cell transformation and tumor development is one of the consequences of aging of genome DNAs. In malignant tumor cells, accumulation of DNA mutations is accelerated by the deficiency of DNA repair and by the rapid cell proliferation.

## 2.3 Accumulation of Misrepairs on cellular level

On cellular level, Misrepairs may take place mainly on the structure of cytoskeleton. A Misrepair appears as an alteration of the organization of some filaments after repair of an injury. Repetition of such altered remodeling of filaments will gradually deform the cytoskeleton of a cell. Deformation of cytoskeleton will lead to a change of the distribution of organelles and a change of the shape of cell and nucleus. This is the reason why aged cells have shrunken cell membrane, deformed nucleus, and agglomerated euchromatins. Cytoskeleton filaments also compose the "roads" for substance transportation in a cell: via filament cables or via cytoplasm flow. Altered remodeling of filaments will affect the intracellular substance-transportation and information-transmission, and reduce the efficiency of cell metabolism and cell adaptation. Thus, accumulation of Misrepairs of cytoskeleton results in aging of the cell. Neurofibrillary tangle is an aging change of neuron cells, in which the neurofibrils are tangling together (Lee, 2005). In our view, neurofibrillary tangle develop from the accumulation of altered remodeling of neurofibrils, due to many times of repair of broken neurofibrils in the neuron cell.

An aged cell has one of the four destinations: **A**. death and replaced by a new cell; **B.** death and replaced by other types of cells/ECMs such as fibroblasts and collagen fibers; **C**. further aging with accumulation of lipofuscin bodies; and **D**. death and isolated by connective tissue when the dead substance is un-degradable. Many changes can be observed in an aged cell, but some of them including the accumulation of lipofuscin bodies are in fact the consequence, but not the cause, of aging of the cell.

## 2.4  Accumulation of Misrepairs on tissue level

A Misrepair on tissue level may appear as a change of the type, the amount, and/or the distribution of cells/ECMs. Accumulation of Misrepairs in a tissue will distort gradually the organization of cells/ECMs. For example, liver cirrhosis is a result of accumulation of Misrepairs of the liver tissue triggered by frequent deaths of hepatocytes in chronic inflammations.  A normal liver has a regular organization of hepatocytes, which are grouped into hepatic lobules. Hepatocytes may die from toxic substances or viral-attacks. The liver can regenerate hepatocytes in case of cell death. However, frequent exposure to toxic substances or viral will make the death of hepatocytes occur repeatedly. The repetition of cell death and the death of many cells in the same time will make the full repair impossible. Some dead hepatocytes have to be replaced by fibroblasts and collagen fibers in repair. The accumulation of collagen fibers, which are in an irregular organization, will alter the structure of liver tissue and makes the liver stiffer.  Liver cirrhosis is a rapid process of aging of the liver, which is accelerated often by viral infections or alcohol-poisoning. Other types of aging changes in tissues including tissue degeneration and irreversible hyperplasia are also results of accumulation of Misrepairs on tissue level.

### A.  *Tissue degeneration*

Atherosclerotic plaques and hyaline degeneration are two typical aging-related degeneration changes. A common change in tissue degeneration is the deposition of abnormal substance in a tissue. The deposed substance can have two sources: un-degradable substance or cell products for repair. Un-degradable substance can be from dead cells or from foreign substance such as dead bacterial. When a large number of cells die in the same time in a local area, some of them cannot be removed completely. In such situation, an adaptive response of the tissue is to isolate them in situ into a fibrotic capsule for rebuilding the structural integrity. Such a repair is a Misrepair. For example, the development of atherosclerotic plaques in arterial walls is a result of accumulation of Misrepairs of endothelium. When part of the endothelium is injured, the lipid in bloodstream can infuse into sub-endothelium. Too much lipid cannot be removed by the local macrophages and they deposit beneath the endothelium. Deposition of lipids will make the complete repair of the endothelium impossible. A fibrotic capsule made by collagen fibers and myofibers is thus developed for isolating the lipids and for sealing the endothelium. The fibrotic capsule and the inner lipids compose into an atherosclerotic plaque. Repeated injuries and repeated lipid infusions in a local part of endothelium results in the gradual enlargement of a plaque.

Another source of deposit substance is the cell products for repair. Like that in cell metabolism and in tissue immunoreactions, all of the activated molecules involved in bio-reactions will be removed immediately after functioning. However, one kind of molecules is exceptional: the proteins for repair. These proteins are used for reconstructing a structure, and they are finally integrated into the remodeled structure. Therefore, the repairing proteins can deposit in tissues permanently. However, in severe injuries, collagen fibers are often used as repairing proteins for replacing dead cells or injured ECMs, and this is Misrepair. For example, in un-regenerable tissues such as skeleton muscle, dead muscular cells are often replaced by collagen fibers. In regenerable tissues such as liver tissue, the tissue lesion due to death of many cells is often closed by collagen fibers. The elastic fibers and myofibers in elastic organs such as the skin and the arterial walls have low potential of regeneration, and the broken fibers are often replaced by collagen fibers. Accumulation of the repairing collagen fibers then leads to the hyaline degeneration in different aging changes, such as atrophy of muscular tissues, arteriosclerosis, and liver cirrhosis.

### B. Irreversible hyperplasia

Hyperplasia is the phenomenon that a tissue produces excessive cells for adaptation. There are three types of hyperplasia: physiologic hyperplasia, pathological hyperplasia, and neoplasia (tumor). Physiologic hyperplasia is a temporary tissue response to an environment change. The cell proliferation will be stopped and the cell number will be reduced to normal level by withdrawing of the stimulators. Therefore, physiologic hyperplasia is reversible. The breast hyperplasia during breeding is a type of physiologic hyperplasia controlled by the levels of hormones. Differently, pathological hyperplasia is irreversible. For example, the synovial hyperplasia in osteoarthritis is a result of repairs of synovial tissue triggered by repeated injuries. The cells in pathological hyperplasia are produced for repair. These cells are genetically normal, but they are often in a different reorganization from the original one. Therefore, pathological hyperplasia is a result of accumulation of Misrepairs of a tissue. In neoplasia, the cell proliferation is uncontrollable and the cells are genetically abnormal.

Neoplasm is a result of accumulation of DNA mutations in somatic cells. DNA mutations in somatic cell are result of Misrepair of DNA breaks (Bishay, 2001). Pathologic hyperplasia and neoplasia are both results of accumulation of Misrepairs; however, the former is on tissue level whereas the latter is on DNA level.

**2.5 Aging of a multi-cellular organism: on tissue level**

The aging of DNAs, cells and tissues can be independent, but a critical question is whether or not aging of an organism is a result of aging of cells and aging of molecules. In fact, in some aging symptoms, such as synovial hyperplasia and atherosclerotic plaques, there is no change on cells or molecules. A common and consistent change in most aging changes is an irreversible change on the organization of cells/ECMs, which is actually on tissue level. In our view, aging of an organism, which is disease-related, takes place essentially and sufficiently on tissue level.

Full functionality of a tissue/organ depends not only on the full functionality of each cell and each ECM, but also on a perfect collaboration (spatial relationship) of them. An altered reorganization of normal cells/ECMs is sufficient for causing a decline of tissue functionality and body functionality; and this change is neither on cellular level nor on molecular level but on tissue level. When dead cells and injured ECMs are replaced by new cells/ECMs of the same type however in an altered reorganization, tissue functionality will be reduced (Figure 3A). When an aged cell remains in a tissue, it alters also the spatial relationship of local cells/ECMs and affects tissue functionality (Figure 3B). Therefore an essential change of aging of an organism, by which we may develop diseases, is an irreversible change of the spatial relationship between cells/ECMs in a tissue. Aging of an organism does not essentially require aging of cells and aging of molecules but requires aging of tissues. Interestingly, in regenerable tissues, deposition of aged cells is an effect rather than a cause of aging of a tissue. On one hand, aging of a tissue can accelerate aging of cells, since the efficiency of substance transportation and cell communication is reduced in an aged tissue. On the other hand, aged cells cannot be efficiently removed, since the repair-efficiency of an aged tissue is also reduced.

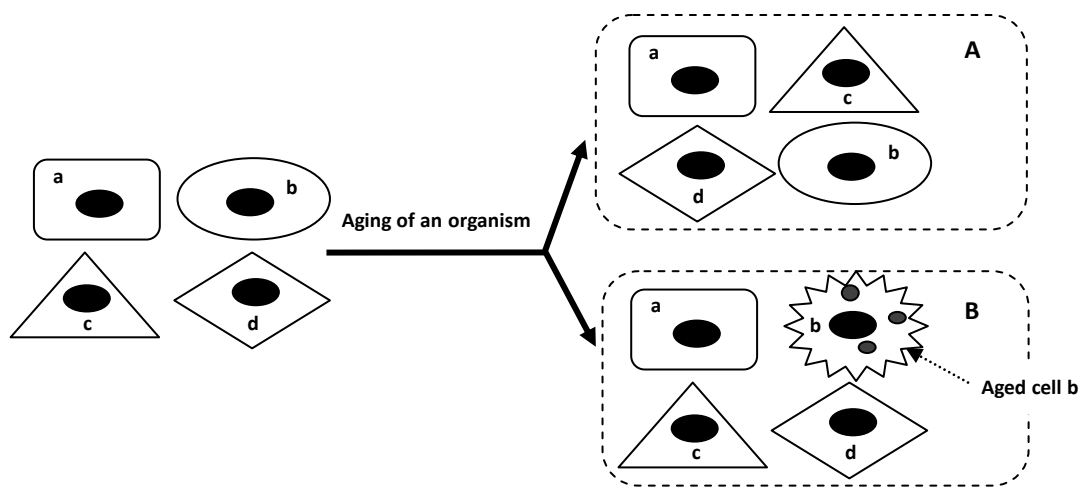

**Figure 3. An essential change of aging of an organism: a change of spatial relationship between cells/ECMs**

An altered reorganization of normal cells/ECMs is sufficient for causing a decline of tissue functionality and body functionality; and this change is neither on cellular level nor on molecular level but on tissue level. When dead cells and injured ECMs are replaced by new cells/ECMs of the same types however in an altered organization (**a, b, c,** and **d**), tissue functionality will be reduced (**A**). When an aged cell remains in a tissue (**aged cell b, in B**), it alters also the spatial relationship of local cells/ECMs and affects tissue functionality (**B**). Therefore an essential change of aging of an organism is an irreversible change of spatial relationship between cells/ECMs in a tissue. Aging of an organism does not essentially require aging of cells and aging of molecules but requires aging of tissues.

### III. Conclusion: Misrepair-accumulation theory

With a generalized concept of Misrepair, we have explained the roles of structure, damage, injury, and repair/maintenance function in the process of aging. By causing injuries, damage promotes the occurrence of Misrepairs, in which damage is the causing factor whereas Misrepair is the final effect. Misrepair is a strategy of repair, and this is essential for preventing death of an organism in cases of severe injuries and repeated injuries. Misrepair mechanism tells us that an organism will make all its efforts to survive in destructive environment. Misrepair mechanism is essential for the survival of an organism till reproduction age; therefore it is essential for the survival of a species. However, Misrepairs alter a living structure permanently, thus they accumulate with time. Accumulation of Misrepairs alters gradually the structure and the functionality of a molecule, a cell and a tissue, appearing as aging of them. Misrepair is therefore the common change underlying all aging changes. Aging can take place on the levels of molecule, cell and tissue, respectively; however aging of an organism takes place essentially and sufficiently on tissue level. We summarize our idea by the Misrepair-accumulation theory: aging of an organism is a result of accumulation of Misrepairs on tissue level.

With this theory, we can understand why we age and how we age. We age because we have to make Misrepairs for surviving when we suffer injuries. We are aging because of the accumulation of Misrepairs of tissues/organs with time. We cannot avoid aging, because without Misrepairs we could not survive till mature age to have children and the species of human being would die out. Aging of individuals is the price to be paid for the survival of the species! Misrepair mechanism reveals that an organism is not programmed to die but to survive as long as possible (Kirkwood, 2005).